\documentclass[two            column,showpacs,preprintnumbers,amsmath,amssymb]{revtex4}

\def\tt{{\tilde t}}
\def\ts{{\tilde s}}
\def\tN{{\tilde N}}
\def\tK{{\tilde K}}

\def\be{\begin{equation}}
\def\ee{\end{equation}}

\def\be{\begin{equation}}
\def\ee{\end{equation}}

\def\be{\begin{equation}}
\def\ee{\end{equation}}

\def\lmk{\left(}
\def\rmk{\right)}
\usepackage[dvipdfmx]{graphicx}
\usepackage{here}
\usepackage{comment}
\usepackage{epsf}
\usepackage{amsmath}
\usepackage{graphics}
\usepackage{amsfonts}
\usepackage{amssymb}
\usepackage{latexsym}
\usepackage{color}
\input{colordvi.tex}
\usepackage{feynmp}

\begin{document}
\preprint{RESCEU-14/16}

\title{Entropic interpretation of the Hawking-Moss bounce}

\author{Naritaka Oshita$^{1,2}$}
\author{Jun'ichi Yokoyama$^{1,2,3}$}
\affiliation{
  $^1$Research Center for the Early Universe (RESCEU), Graduate School
  of Science,\\ The University of Tokyo, Tokyo 113-0033, Japan
}
\affiliation{
  $^2$Department of Physics, Graduate School of Science,\\ The University of Tokyo, Tokyo 113-0033, Japan
}
\affiliation{
  $^3$Kavli Institute for the Physics and Mathematics 
of the Universe (Kavli IPMU), WPI, UTIAS,\\
The University of Tokyo, Kashiwa, Chiba 277-8568, Japan
}

\date{\today}

\begin{abstract}
We revisit the derivation of the Hawking-Moss transition rate.  Using the static coordinates, we show that the Euclidean action is
entirely determined by the contribution of the entropy of de Sitter
 space
which is proportional to the surface area of the horizon.
This holographic feature is common to any static spacetime with a horizon on which the shift vector vanishes.
\end{abstract}


\maketitle

When the Hawking-Moss bounce was first discovered \cite{Hawking:1981fz}, it was interpreted as describing quantum
tunneling from a de Sitter universe as a whole to another de Sitter space with a larger (effective) cosmological constant.
Since a transition to a state with larger energy density is
counterintuitive,
 many people perceived it with surprise.
Another counterintuitive aspect
is that it depends only on the potential energy densities before and after the transition independent of the ``distance''
in the field space.

The latter point was challenged by Weinberg later \cite{Weinberg:2006pc}, who proposed a ``thermal'' interpretation in the limit the difference of the vacuum energy densities is small so that geometrically both states are described by de Sitter space with the small Hubble parameter.

Nowadays the Hawking-Moss instanton is playing more and more important roles in various fields of physics---not
only in the inflationary cosmology \cite{Sato:2015dga}, where it was originally applied, but also in the landscape of
string theory \cite{Susskind:2003kw} where exponentially large number of possible vacuum states
with different  energy densities exist.

In this \textit{Letter},  we revisit the interpretation of the Hawking-Moss instanton to provide a new picture, namely, 
the entropic interpretation without any restrictions unlike in \cite{Weinberg:2006pc}.
The importance of the gravitational entropy term in the vacuum transition rate has been stressed by 
Gregory et al.\ in a different context \cite{Gregory:2013hja}, who incorporated the effect of black hole on the false vacuum decay and showed that a
term proportional to the black hole horizon area must be taken into account.

Before embarking on the entropic interpretation, we review
the derivation of the Hawking-Moss bounce solution \cite{Hawking:1981fz,Weinberg:2006pc,Vilenkin:1983xq}.
Assuming that the bounce solution has O$(4)$ symmetry, the Euclidean metric can be characterized by one parameter
$\xi$ and its function $\rho(\xi)$,
\begin{eqnarray}
ds^2 = d\xi^2 + \rho(\xi)^2 d\Omega^2_{\text{III}},
\end{eqnarray}
where $d \Omega_{\text{III}}$ represents the line element of the unit three-sphere.
The Euclidean action of the Einstein gravity and a canonical scalar field $\phi$ with a potential $V(\phi)$ is then written as follows.
\begin{eqnarray}
I_{\text{E}} \!=\!
 2 \pi^2 \!\!\int\!\! d\xi \left[ \rho^3\! \left( \frac{1}{2}
					    \dot{\phi}^2\! +\! V(\phi)\! \right)\! +\!
\frac{3}{8 \pi G} \left( \rho^2 \ddot{\rho} +\! \rho \dot{\rho}^2\! -\! \rho \right) \right]
\label{011901}
\end{eqnarray}
where an over-dot represents differentiation with respect to $\xi$.
From (\ref{011901}), the field equations read,
\begin{eqnarray}
&&\ddot{\phi} + \frac{3 \dot{\rho}}{\rho} \dot{\phi} = \frac{d V}{d \phi}, \label{011902} \\
&&\dot{\rho}^2 = 1+ \frac{8 \pi G}{3} \rho^2 \left( \frac{1}{2} \dot{\phi}^2 -V \right). \label{011903}
\end{eqnarray}
The Hawking-Moss solution corresponds to a static scalar field configuration with $\dot{\phi} = \ddot{\phi} = 0$
which is realized at potential extrema with $dV / d \phi = 0$. Hence (\ref{011903}) reads
\begin{eqnarray}
\dot{\rho}^2 = 1- \frac{8\pi G}{3} \rho^2 V \label{011904}
\end{eqnarray}
and its solution is given as
\begin{eqnarray}
&&\rho( \xi) = H_{\text{s}}^{-1} \sin{\left( H_{\text{s}} \xi \right)},\label{011905} \\
&&H_{\text{s}}^2 \equiv \frac{8 \pi G}{3} V(\phi_{\text{s}}),
\end{eqnarray}
where $\phi_{\text{s}}$ is a field value at a potential extremum.
Substituting the solution (\ref{011905}) into the action (\ref{011901}),
we find
\begin{eqnarray}
I_{\text{E}}(\phi_{\text{s}}) = -\frac{3}{8G^2 V(\phi_{\text{s}})}.
\end{eqnarray}
Hawking and Moss \cite{Hawking:1981fz} originally considered the transition from a false vacuum state $\phi_{\text{s}} = \phi_{\text{fv}}$
to the local potential maximum $\phi_{\text{s}} = \phi_{\text{top}}$ and identified the transition rate as
\begin{eqnarray}
&&\Gamma_{\text{fv} \to \text{top}} = A e^{-B_{\text{HM}}} 
= A \exp{\left[ -I_{\text{E}}(\phi_{\text{top}}) + I_{\text{E}} (\phi_{\text{fv}}) \right]} \nonumber \\
&&~~~~~~~~~~=A \exp{\left[ \frac{3}{8G^2} \left( \frac{1}{V(\phi_{\text{top}})} - \frac{1}{V(\phi_{\text{fv}})} \right) \right]}, \label{rate}
\end{eqnarray}
where the prefactor $A$ may be estimated as $A \sim H^4(\phi_{\text{fv}})$ on dimensional grounds.

In \cite{Weinberg:2006pc}, 
Weinberg proposed a thermal interpretation assuming $\Delta V/V(\phi_{\text{fv}}) \equiv \left[ V(\phi_{\text{top}}) - V(\phi_{\text{fv}}) \right]/V(\phi_{\text{fv}}) \ll 1$, when $B_{\text{HM}}$ is given by
\begin{eqnarray}
&&B_{\text{HM}} \simeq \frac{\Delta E}{T_{\text{H}}} \label{020301} \\
&&\text{with} \ \Delta E = \frac{4 \pi}{3} H^{-3} (\phi_{\text{fv}}) \Delta V
 \ \text{and} \ T_{\text{H}} = \frac{H (\phi_{\text{fv}})}{2 \pi}. \nonumber
\end{eqnarray}
Here $\Delta E$ is the potential energy increment in the horizon $H^{-1} (\phi_{\text{fv}}) (\simeq H^{-1} (\phi_{\text{top}}))$,
and $T_{\text{H}}$ is the Hawking temperature of de Sitter space. He argues  that the gravitational effect is negligible
because the geometry does not change practically 
before and after the transition 
thanks to the assumption $\Delta V/V \ll 1$.
As a result the formula based on (\ref{020301})  is identical to the case a horizon-sized domain receives thermal fluctuation at the
Hawking temperature.

In the rest of this \textit{Letter}, however, we show that the exponent $B_{\text{HM}}$ is completely determined by
the gravitational entropy of the system, as the bulk energy
of the scalar field is fully canceled out by the negative gravitational energy due to the Hamiltonian constraint.
It is therefore concluded that only the gravitational entropy affects the Hawking-Moss
transition, and that it does not  break the conservation
of energy.

In order to prove the above statement, it is essential to describe the (Euclidean) de Sitter space with a static metric $({\mathcal M},g_{\mu \nu})$.
Here, we start with a more general Arnowitt-Deser-Misner (ADM) decomposition \cite{Arnowitt:1959ah}
\begin{equation}
 ds^2=-N^2 dt^2 + h_{ij} (dx^i + N^i dt) (dx^j + N^j dt) \label{Lorentzian}
\end{equation}
where $N$ is the lapse function, $N^i$ is the
shift vector, and $h_{ij}$ is the spatial metric.  The Latin indices run from $1$ to $3$.
Applying the Wick rotation $t = -i\tt$ to introduce the Euclidean time
$\tt$, (\ref{Lorentzian}) reads
\begin{eqnarray}
d\ts^2 =N^2 d \tt^2 + h_{ij} (dx^i + \tN^i d\tt) (dx^j +\tN^j d \tt)
\end{eqnarray}
with $\tN^i \equiv -iN^i$.  Here and hereafter we put a tilde on
quantities in the Euclidean space which is multiplied by some power of
$i$ upon Wick rotation.  
Correspondence to the unrotated Lorentzian
counterpart is also shown below.  For example, the extrinsic curvature of the
$\tt=$const.\ three-space $\Sigma_{\tt}$ is expressed as
\begin{equation}
\tK_{ij}=\frac{1}{2N}\lmk \frac{\partial h_{ij}}{\partial \tt}
-D_i\tN_j-D_j\tN_i\rmk = -iK_{ij},
\end{equation}
where $D_i$ denotes covariant derivative with respect to $h_{ij}$.

The Euclidean Einstein action $I_{\text{E}}^{\text{(G)}}$ is expressed as
\begin{widetext}
\begin{eqnarray}
&&I_{\text{E}}^{\text{(G)}}
= - \frac{1}{16 \pi G} \int_{{\mathcal M}} d^3 xd\tt  \sqrt{\tilde g} {\tilde R} \nonumber \\
&&~~~~~~=\int d \tt \left[ \int_{\Sigma_{\tt}} d^3 x \left( \tilde{\pi}^{ij} \partial_{\tt} h_{ij}
+ N \tilde{\mathcal H}^{\text{(G)}} - \tilde{N}^i \tilde{\mathcal H}^{\text{(G)}}_i \right) - \int_{\text{S}} d^2 x \sqrt{\sigma} 
\left( \frac{n^i \partial_i N}{8 \pi G} - \frac{2}{\sqrt{h}} n_i \tilde{N}_j \tilde{\pi}^{ij} \right) \right].
\label{020302}
\end{eqnarray}
Here $\tilde{\pi}^{ij}$ is the Euclidean momentum conjugate to $h_{ij}$, $\sigma_{ij}$ is an induced metric on the boundary surface
S with $\sigma \equiv \text{det} \sigma_{ij}$,
and $n^i$ is the unit normal vector on the boundary surface S where we assume $\partial_i \tilde{N}_j$ vanishes.
$\tilde{\mathcal H}^{\text{(G)}}$ and $\tilde{\mathcal H}_i^{\text{(G)}}$ are the gravitational Hamiltonian and the momentum for the dynamics of the foliation $\Sigma_{\tt}$.
They are given by 
\begin{eqnarray}
\tilde{\mathcal H}^{\text{(G)}} = \frac{\sqrt{h}}{16 \pi G} \left(- {}^{(3)}R - \tilde{K}_{ij} \tilde{K}^{ij} + \tilde{K}^2 \right),\ \ \ \ 
\tilde{\mathcal H}^{\text{(G)}}_i = -2 h_{ij} D_k \tilde{\pi}^{jk},\ \ \ \ 
\tilde{\mathcal \pi}^{ij} = \frac{\sqrt{h}}{16 \pi G} \left( \tilde{K}^{ij} - h^{ij} \tilde{K} \right)=-i \pi^{ij},
\label{012101}
\end{eqnarray}
where ${}^{\tiny{(3)}} R$  denotes the three-curvature   on the hypersurface $\Sigma_{\tt}$ and $\tilde{K}$ represents the trace of the extrinsic curvature $h^{ij} \tilde{K}_{ij}$.

The matter Euclidean action, on the other hand, is expressed as
\begin{eqnarray}
&&I_{\text{E}}^{\text{(M)}} = \int d^3 xd\tt \sqrt{\tilde g} \left[ 
\frac{1}{2} {\tilde g}^{\mu \nu} \partial_{\mu} \phi \partial_{\nu} \phi
+ V(\phi) \right]
 = \int d{\tt} \int_{\Sigma_{\tt}} d^3x \left(  \tilde{\mathcal P}_{\phi} \partial_{\tt} \phi
+N \tilde{\mathcal H}^{\text{(M)}} - \tilde{N}^i \tilde{\mathcal H}_i^{\text{(M)}} \right)
\end{eqnarray}
where 
\begin{equation}
\tilde{\mathcal P}_{\phi} = \sqrt{h}\left[\frac{1}{N} \partial_{{\tt}}\phi - \frac{\tilde{N}^i}{N} \partial_i \phi\right]
\end{equation}
 is the momentum conjugate to $\phi$, and 
 \begin{align}
 \tilde{\mathcal H}^{\text{(M)}}& =\sqrt{h}\left[- \frac{1}{2} \left( \frac{1}{N}\partial_{\tt}\phi -\frac{\tilde{N}^i}{N} \partial_i \phi \right)^2 + \frac{1}{2}h^{ij} \partial_i \phi
\partial_j \phi + V\right], \ \ \ \ 
\tilde{\mathcal H}^{\text{(M)}}_i =  \sqrt{h} \left[\frac{1}{N} \partial_{{\tt}} \phi\partial_i \phi-\frac{\tilde{N}^j}{N} \partial_j \phi \partial_i \phi\right]
\end{align}
\end{widetext}
 are the matter part of the Hamiltonian and momentum, respectively. 
 
Classical Euclidean solutions are found by taking variation of the total Euclidean action $I_{\text{E}}^{\text{(tot)}}
= I_{\text{E}}^{\text{(G)}} + I_{\text{E}}^{\text{(M)}}$.
From variation with respect to $N$ and $\tN^i$, we find 
the Hamiltonian and the momentum constraints,
\begin{align}
\tilde{\mathcal H}^{\text{(tot)}} \equiv \tilde{\mathcal H}^{\text{(G)}} +
\tilde{\mathcal H}^{\text{(M)}} = & 0, \ \ \ 
\tilde{\mathcal H}^{\text{(tot)}}_i \equiv \tilde{\mathcal H}^{\text{(G)}}_i +
\tilde{\mathcal H}^{\text{(M)}}_i = 0. 
\end{align}
Therefore for a static configuration with $\partial_{\tt} h_{ij} = 0$
and $\partial_{\tt} \phi = 0$, the total Euclidean action is simply given by
the surface terms as
\begin{eqnarray}
I_{\text{E static}}^{\text{(tot)}} = - \int_{\text{S}} d \tt d^2 x \sqrt{\sigma} \left( \frac{n^i \partial_{i} N}{8 \pi G} - \frac{2}{\sqrt{h}} n_i \tilde{N}_j \tilde{\pi}^{ij} \right)  \!\! .
\label{AA}
\end{eqnarray}

For the particular case of de Sitter space, the static metric is given by
\begin{eqnarray}
&&d\ts^2=\tilde{g}_{\mu \nu} dx^{\mu} dx^{\nu} \nonumber \\
&&~~~~= \left( 1-H^2 r^2 \right) d\tt^2 + \frac{dr^2}{ 1-H^2 r^2} +r^2 d\Omega_{\text{II}}^2, 
\end{eqnarray}
where $H$ is the Hubble parameter, $r$ denotes radial coordinate and $d \Omega_{\text{II}}^2$ is the metric on the unit two sphere.
In the following, we impose the periodic boundary condition on the Euclidean time $\tt$ with a period $\beta$.

Introducing a foliation $\Sigma_{\tt}$ in the spacetime fixed at constant
Euclidean time $\tt$, which takes the value in the range $0 \leq \tt< \beta$,
we can easily decompose the Euclidean de Sitter metric with
\begin{align}
N =&  \sqrt{1-H^2 r^2}, \ \tN^i = 0, \label{012902} \\
h_{\mu \nu} =&   g_{\mu \nu} - t_{\mu} t_{\nu} \nonumber \\
 =&   \text{diag} \left(0,(1-H^2r^2)^{-1}, \ r^2, \ r^2 \sin^2{\theta} \right) \label{012901}\\
t_{\mu} \equiv&   (\sqrt{1-H^2 r^2},0,0,0),
\label{012002}
\end{align}
where $t_{\mu}$ is the unit normal vector on the hypersurface $\Sigma_{\tt}$.
This manifold generally has a conical singularity at $r = 1/H$ where $N = 0$. This implies
that the curvature is divergent on the de Sitter horizon, although the horizon
is not a physical singularity.  

As we see below, the conical singularity can be avoided by a specific choice of $\beta$, namely the inverse Hawking temperature $\beta_{\text{H}} \equiv 2\pi /H$.
It should be noted, however, that the manifold still collapses to a single point
on the horizon and,
as shown below, this plays an important role in deriving the entropy term from the Euclidean action.

\begin{figure}[h]
  \begin{center}
    \includegraphics[keepaspectratio=true,height=54mm]{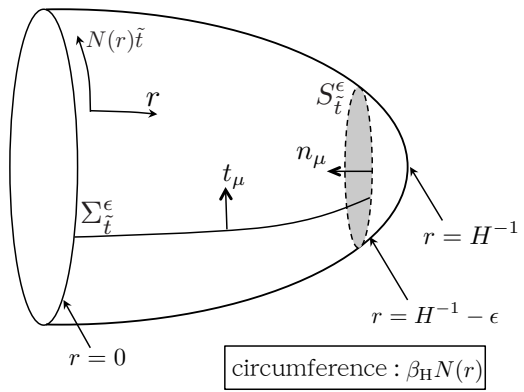}
  \end{center}
  \caption{
  The manifold $\mathcal M$ including the horizon
  at $r= 1/H$ where $N = 0$. We can regularize the Euclidean action by introducing 
  a hypothetical boundary at $r=H^{-1}-\epsilon$ denoted by $\text{S}=S^{\epsilon}_{\tt}$ with the
   cut off parameter
  $\epsilon$ set to zero at the end of the calculation.
  }%
  \label{012001}
\end{figure}

In the following, therefore, we regularize the collapsing part of the manifold by first
restricting 
the integration to the region ${\mathcal M}_{\epsilon} \equiv \left\{ x^{\mu} : r \leq \frac{1}{H} - \epsilon \right\}$
and then setting the regularization parameter $\epsilon$ to zero after the calculation of
the Euclidean action (Fig. \ref{012001}).

In this case the action (\ref{AA}) has only the first term, where the surface $S_{\tt}^{\epsilon}$ is located at
$r = 1/H - \epsilon$ with its normal vector $n_{\mu}$ given by 
\begin{eqnarray}
n_{\mu} = (0,-1/\sqrt{1-H^2 r^2}, 0, 0).
\end{eqnarray}
Note that, since this surface is not a real boundary of the theory, being introduced 
just for the sake of regularization, one should not apply the Gibbons-Hawking boundary terms \cite{GH,GHb} here.
The length of circumference of the manifold $\mathcal M$ is $\beta N$
and the relation 
\begin{eqnarray}
\displaystyle \lim_{\epsilon \to 0} n^i \partial_i \left[ \beta N \right] =\beta H= 2 \pi
\end{eqnarray}
should be satisfied to
ensure the absence of the conical singularity.  This is the reason we must identify $\beta$ with  the inverse Hawking temperature $\beta_{H}$.

Hence, the nonvanishing term in (\ref{AA}) is calculated as
\begin{eqnarray}
\displaystyle \lim_{\epsilon \to 0}\left[ - \beta_{H} \int_{S_{\tt}^{\epsilon}} d^2 x \sqrt{\sigma}
 \frac{n^i \partial_{i} N}{8 \pi G} \right]
 = - \frac{A}{4 G},
\end{eqnarray}
where $A$ is the area of the horizon given as
\begin{eqnarray}
\left. A \equiv \int_{S_{\tt}^{\epsilon}} d^2 x \sqrt{\sigma} \right|_{\epsilon = 0}  = \frac{4 \pi}{H^2}.
\end{eqnarray}

Thus, for the case $\phi = \phi_{\text{s}}$ giving a static de Sitter space with potential energy density $V(\phi_{\text{s}})$,
the action of the Hawking-Moss instanton 
\begin{eqnarray}
 I_{\text{E}}^{\text{(tot)}} (\phi_{\text{s}}) = - \frac{A(\phi_{\text{s}})}{4 G} = \frac{\pi}{G H^2} = \frac{3}{8 G^2 V(\phi_{\text{s}})}
\end{eqnarray}
is entirely given by the contribution of the de Sitter entropy \cite{GH}.  This is primarily because in the static configuration
the bulk term of the action vanishes due to the Hamiltonian constraint.

In this sense, one can extend the thermal interpretation of the Hawking-Moss solution more rigorously to argue that $e^{-I_{\text{E}}(\phi_s)}$ is indeed proportional to the thermodynamical probability of the state $\phi=\phi_s$, $e^{-F/T}$, 
 where $F = E-TS$ is the free energy.
Here, since $E = 0$, $e^{-F/T}$ simply reads $e^{S} = e^{\ln W(\phi_{\text{s}})} = W(\phi_{\text{s}})$.
In other words, the probability is just proportional to the number of internal states $W(\phi_{\text{s}})$ associated with the de Sitter
space with the energy density $V(\phi_{\text{s}})$.
Thus the smallness of the transition rate (\ref{rate}) to a state with a higher potential energy density 
is not due to the largeness of the energy---in fact, the total energy is always zero---, but  
because of the smallness of the number of microscopic states there.

It is also interesting to note that the Hawking-Moss transition apparently violates the second law of thermodynamics, as it is a transition to a state with  {\it smaller} entropy.
However, since it is governed by a single order parameter $\phi$, the transition itself is 
not a  macroscopic process but a mere microscopic process analogous to the Brownian motion.  Indeed the stochastic approach of inflation can also reproduce the desired probability distribution \cite{Linde:1991sk}.

In conclusion, by using a static coordinate system, we have shown that the
Hawking-Moss instanton is entirely given by the entropy of the de Sitter spacetime
proportional to the horizon area.
In this sense, this solution is  holographic \cite{'tHooft:1993gx}.  
As is clear from the above analysis, 
this feature is common to any static spacetime with a horizon on which the shift vector vanishes $N^i = 0$.   In our derivation,
the Hamiltonian constraint, which asserts that the sum of the material energy and gravitational energy vanishes,  plays an important role.  Hence previous considerations 
based only on the energy of matter part \cite{Weinberg:2006pc,HenryTye:2008xu} cannot grasp the essential feature of the problem.

\textit{Acknowledgements.}
We are grateful to Masafumi Fukuma and Yuho Sakatani for valuable communications.
This work was partially supported by JSPS Grant-in-Aid for Scientific Research
15H02082 (J.Y.), Grant-in-Aid for Scientific Research on Innovative Areas No.
15H05888 (J.Y.), and a research
program of the Advanced Leading Graduate Course for
Photon Science (ALPS) at the University of Tokyo (N.O.).


\end{document}